\documentclass[prd,twocolumn,showpacs,floatfix,amsmath,amssymb,floatfix]{revtex4}
\usepackage{graphicx,dcolumn,booktabs,bm}
\usepackage{longtable,lscape,multirow}
\usepackage{txfonts}
\usepackage{amssymb}
\usepackage{indentfirst}
\usepackage{color,ulem}

\usepackage{graphicx}
\usepackage{epstopdf}
\usepackage{subfigure}

\graphicspath{{Figures/}} 

\newcounter{RomanNumber}

\begin{document}


\title{Predicting exotic molecular states composed of nucleon and $P$-wave charmed meson}
\author{Rui Chen$^{1,2}$}\email{chenr2012@lzu.edu.cn}
\author{Zhi-Feng Sun$^{1,2}$}\email{sunzhif09@lzu.edu.cn}
\author{Xiang Liu$^{1,2}$\footnote{Corresponding author}}\email{xiangliu@lzu.edu.cn}
\affiliation{$^1$Research Center for Hadron and CSR Physics,
Lanzhou University $\&$ Institute of Modern Physics of CAS,
Lanzhou 730000, China\\
$^2$School of Physical Science and Technology, Lanzhou University,
Lanzhou 730000, China}
\author{S. M. Gerasyuta}\email{gerasyuta@SG6488.spb.edu}
\affiliation{Department of Physics, St. Petersburg State Forest Technical University, Institutski Per. 5, St. Petersburg 194021, Russia}

\date{\today}

\begin{abstract}
In this work, we study the interaction between a nucleon and a $P$-wave charmed meson in the $T$ doublet by exchanging a pion.
Our calculations indicate that a nucleon and a $P$-wave charmed meson with $J^P=0^+$ or $J^P=1^+$ in the $T$ doublet
can form bound states. We propose the experimental search for these exotic molecular states near the $D_1(2420)N$ and $D_2^*(2460)N$ thresholds, where Belle, LHCb and the forthcoming Belle II have the discovery potential for them.
\end{abstract}

\maketitle

\section{Introduction}\label{sec1}

Searching for the exotic states is an important and intriguing research topic which can deepen our understanding of the nonperturbative behavior of quantum chromodynamics (QCD). Exotic states have the special configuration beyond the conventional meson and baryon, which include glueball, hybrid, tetraquark state, molecular state, and so on. Among these listed exotic states, molecular state is one of the most popular ones since there were extensive discussions of whether the observed charmonium-like states $XYZ$ can be explained under the molecule configuration (see a recent review for more details \cite{Liu:2013waa}).

\begin{figure}
\includegraphics[scale=0.51]{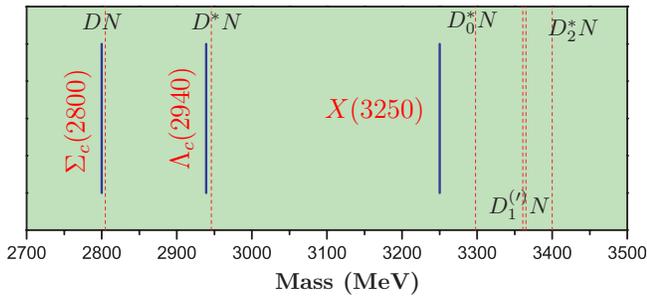}
\caption{The comparison of the masses of $\Lambda_c(2940)$, $\Sigma_c(2800)$, and $X(3250)$ with the $DN$, $D^*N$, $D_0^*N$, $D_1^{(\prime)}N$, and $D_2^*N$ thresholds.}\label{fig:feyn}
\end{figure}

Besides discussing these observed charmonium-like states as the molecular states,
there were some explorations of newly observed charmed hadrons $\Lambda_c(2940)$, $\Sigma_c(2800)$, and $X(3250)$ under the molecule assignments.
$\Lambda_c(2940)$ was first reported by the BaBar Collaboration in the $D^*p$ invariant mass spectrum, where its mass and width are $m=2939.8\pm1.3(\mathrm{stat})\pm1.0(\mathrm{syst})$ MeV and $\Gamma=17.5\pm5.2(\mathrm{stat})\pm5.9(\mathrm{syst})$ MeV \cite{Aubert:2006sp}, respectively. Later, the Belle's analysis of the $\Sigma_c(2455)^{0,++}\pi^{+,-}$ channels also indicates the existence of $\Lambda_c(2940)$ \cite{Abe:2006rz}. A typical property of $\Lambda_c(2940)$ is that $\Lambda_c(2940)$ is near the threshold of $D^*p$, which stimulates the theorist's interest in studying the $D^*N$ interaction and the corresponding  $D^*N$ molecular state \cite{He:2006is,He:2010zq,Ortega:2012cx}.
An isotriplet $\Sigma_c(2800)$ was observed in its $\Lambda_c\pi$ channel, which favors the spin-parity quantum number $J^P=3/2^-$ \cite{Mizuk:2004yu}. Later, BaBar confirmed the Belle's observation of $\Sigma_c(2800)$ \cite{Aubert:2008ax}. Considering the observed $\Sigma_c(2800)$ near the $DN$ threshold, the authors in Ref. \cite{Dong:2010gu} proposed the $DN$ molecular state explanation to $\Sigma_c(2800)$.
In 2006, Mizutani and Ramos studied the $DN$ interaction by adopting a set of in-medium $DN$
coupled channel equations to be solved self-consistently \cite{Mizutani:2006vq}, where they also reviewed three methods, i.e., the QCD sums rule, the nuclear mean field approach, and the self-consistent coupled channels method, which are usually applied to study the behavior of the $D$ meson in nuclear matter (see Ref. \cite{Mizutani:2006vq} for a detailed review).  In Ref. \cite{GarciaRecio:2008dp}, the $S$-wave charmed baryon resonances were investigated by the coupled-channel approach with heavy quark symmetry, where their result indicates that $\Lambda_c(2595)$
can be a $ND^*$ quasi-bound state in the SU(8) symmetry \cite{GarciaRecio:2008dp}, which is different from the conclusion derived from the SU(4) approaches \cite{GarciaRecio:2008dp}.

The remaining $X(3250)$ is a new enhancement structure existing in the $\Sigma_c^{++}\pi^-\pi^-$ invariant mass spectrum \cite{Lees:2012kc} which was studied by BaBar. $X(3250)$ was explained to be a $D_0^*(2400)N$ molecular hadron \cite{He:2012zd,Zhang:2012xx}. In Fig. \ref{fig:feyn}, we compare $\Lambda_c(2940)$, $\Sigma_c(2800)$, and $X(3250)$ with the corresponding thresholds.

By the above introduction, we can form a systematic molecular state picture to describe those observed in $\Lambda_c(2940)$, $\Sigma_c(2800)$, and $X(3250)$, which refers to the interaction between charmed meson and nucleon. In the heavy quark limit, we can categorize charmed mesons into different doublets according to the ${j}_\ell^P$ quantum number, i.e., the $H=(0^-,1^-)$, $S=(0^+,1^+)$, and $T=(1^+,2^+)$ doublets correspond to $j_\ell^P=1/2^-,\,1/2^+$, and $3/2^+$, respectively. At present, the charm meson in the $S$ and $H$ doublets interacting with nucleon was studied when revealing the properties of $\Lambda_c(2940)$, $\Sigma_c(2800)$, and $X(3250)$. Thus, it is natural to focus on these molecular states composed of a nucleon and a $P$-wave charmed meson in the $T$ doublet. Exploring these molecular states can further test the complement of the molecular state family with a nucleon and a charmed meson. In addition, we also notice that the charmed mesons in the $T$ doublet have narrow widths, which shows that these charmed mesons can be act as the suitable component to form the molecular state. For this reason, in this work we perform the dynamical study of the interaction between a nucleon and the charmed meson in the $T$ doublet by the one-pion exchange (OPE) model, which will determine whether molecular states composed of a nucleon and the charmed meson in the $T$ doublet exist. Our prediction of these molecular states can provide valuable information for further experimental search on these molecular states.

This paper is organized as follows.  After the Introduction, we present the detailed deduction of the effective potential depicting the interaction of a nucleon and the charmed meson in the $T$ doublet via one-pion exchange. In Sec. \ref{sec3}, the numerical result will be given. The paper ends with a discussion and a conclusion.

\section{The deduction of effective potential}\label{sec2}

\subsection{The flavor wave function}

Just introduced in Sec. \ref{sec1}, in this work we investigate the interaction of the charmed meson in the $T$ doublet with a nucleon, where the charmed meson in the $T$ doublet is either $D_1$ with $J^P=1^+$ or $D_2^*$ with $J^P=2^+$.
The discussed $D_1N$ and $D_2^*N$ molecular systems can be categorized into several groups in terms of the $I(J)$ quantum numbers, i.e.,
\begin{eqnarray}
D_1N:  \begin{array}{ccccc}
       &|I\left(\frac{1}{2}^{}\right)\rangle:   &|{}^2\mathbb{S}_{\frac{1}{2}}\rangle,   &|{}^4\mathbb{D}_{\frac{1}{2}}\rangle.
\\
&|I\left(\frac{3}{2}^{}\right)\rangle:       &|{}^4\mathbb{S}_{\frac{3}{2}}\rangle,   &|{}^2\mathbb{D}_{\frac{3}{2}}\rangle,
&|{}^4\mathbb{D}_{\frac{3}{2}}\rangle.
\end{array}\label{1}\\
D_2^*N:
 \begin{array}{ccccc} &|I\left(\frac{1}{2}^{}\right)\rangle:   &|{}^4\mathbb{D}_{\frac{1}{2}}\rangle,   &|{}^6\mathbb{D}_{\frac{1}{2}}\rangle,\\
&|I\left(\frac{3}{2}^{}\right)\rangle:   &|{}^4\mathbb{S}_{\frac{3}{2}}\rangle,   &|{}^4\mathbb{D}_{\frac{3}{2}}\rangle,    &|{}^6\mathbb{D}_{\frac{3}{2}}\rangle.\\
&|I\left(\frac{5}{2}^{}\right)\rangle:   &|{}^6\mathbb{S}_{\frac{5}{2}}\rangle,   &|{}^4\mathbb{D}_{\frac{5}{2}}\rangle,    &|{}^6\mathbb{D}_{\frac{5}{2}}\rangle,\\
&|I\left(\frac{7}{2}^{}\right)\rangle:   &|{}^4\mathbb{D}_{\frac{7}{2}}\rangle,   &|{}^6\mathbb{D}_{\frac{7}{2}}\rangle.
\end{array}\label{2}\end{eqnarray}
Here, $I$ and $J$ denote the isospin and the total angular momentum of the molecular system, respectively. $I$ can be taken as 0 or 1. For the $D_1N$ systems, the possible $J$ quantum numbers are $1/2$ and $3/2$, while we have $J=1/2,\,3/2,\,5/2,\, 7/2$ for the $D_2^*N$ systems. In addition, the corresponding component $|^{2S+1}L_J\rangle$ of each system is listed, where $S$ and $L$ denote the total spin and the orbital angular momentum between $D_1/D_2^*$ and $N$
of the corresponding system, respectively. Thus, $L=0$ and $L=2$ are marked by $\mathbb{S}$ and $\mathbb{D}$, respectively.
For the isovector $D_1N$ systems, their flavor wave functions are $|D_1^+p\rangle_f$, $\frac{1}{\sqrt{2}}\left(|D_1^+n\rangle_f-|D_1^0p\rangle_f\right)$, and $|D_1^0n\rangle_f$ with $(I,I_3)=(1,1)$, $(1,0)$, and $(1,-1)$, respectively.
For the isoscalar $D_1N$ system, its flavor wave function is $\frac{1}{\sqrt{2}}\left(|D_1^+n\rangle_f-|D_1^0p\rangle_f\right)$ with $(I,I_3)=(0,0)$. The definition of the flavor wave function of the $D_2^*N$ systems is similar to that of the $D_1 N$ systems.
We notice that there is only $D$-wave contribution to the $D_2^*N$ systems with $J=1/2$ and $J=7/2$. Thus, in this work we discard these two $D_2^*N$ systems since we only consider the $S$-wave interaction between $D_1/D_2^*$ and $N$ and include the $S$-wave and $D$-wave mixing effect.

When discussing the concrete $D_1N$ and $D_2^*N$ systems, we  can write out the general expressions of their wave function, i.e.,
\begin{eqnarray}
|D_1N({}^{2S+1}L_{J})\rangle &=& \sum_{m,m',m_Sm_L}C^{S,m_S}_{\frac{1}{2}m,1m'}C^{J,M}_{Sm_S,Lm_L}
          \chi_{\frac{1}{2}m}\epsilon^{m'}|Y_{L,m_L}\rangle,\label{001}\nonumber\\
|D_2^*N({}^{2S+1}L_{J})\rangle &=& \sum_{m,m'',m_Sm_L}C^{S,m_S}_{\frac{1}{2}m,2m''}C^{J,M}_{Sm_S,Lm_L}
          \chi_{\frac{1}{2}m}\zeta^{m''}|Y_{L,m_L}\rangle,\label{002}\nonumber
\end{eqnarray}
which will be applied to the calculation of the effective potential,
where $\chi_{\frac{1}{2}m}$ denotes the spin wave function. $Y_{L,m_L}$ is defined as the spherical harmonics function, while $C^{J,M}_{Sm_S,Lm_L}$, $C^{S,m_S}_{\frac{1}{2}m,1m'}$ and $C^{S,m_S}_{\frac{1}{2}m,2m''}$ are the Clebsch-Gordan coefficients. $\epsilon^{m'}$ ($m'=0,\pm1$) and $\zeta^{m''}$ ($m''=0,\pm1,\pm2$) are the polarization vectors of $D_1$ and $D_2^*$, respectively, which have definitions
$\epsilon^{\pm 1}=\mp\frac{1}{\sqrt{2}}\left(\epsilon_x\pm i\epsilon_y\right)$ and $\epsilon_0=\epsilon_z$.
The special expressions of the polarization vector for the $D_1$ meson
are $\epsilon^{\pm1}= \frac{1}{\sqrt{2}}\left(0,\pm1,i,0\right)$ and
$\epsilon^{0}= \left(0,0,0,-1\right)$.
Similarly, $\zeta^{m''}$, as the polarization tensor of the $D_2^*$ meson, can be constructed by $\zeta^{m''} = \sum_{m_1,m_2}\langle1,m_1;1,m_2|2,m''\rangle\epsilon^{m_1}\epsilon^{m_2}$, i.e.,
\begin{eqnarray}
\zeta^{\pm2}&=& \epsilon^{\pm1}\epsilon^{\pm1},\nonumber\\
\zeta^{\pm1} &=& \sqrt{\frac{1}{2}}\left(\epsilon^{\pm1}\epsilon^{0}
                              +\epsilon^{0}\epsilon^{\pm1}\right),\nonumber\\
\zeta^{0}&=& \sqrt{\frac{1}{6}}\left(\epsilon^{+1}\epsilon^{-1}
            +\epsilon^{-1}\epsilon^{+1}+2\epsilon^{0}\epsilon^{0}\right).\nonumber
\end{eqnarray}

\subsection{The effective Lagrangian}\label{subsec2}

For obtaining the effective potential of the $D_1N$ and $D_2^*N$ systems, the OPE model is adopted, where the $D_1N$ and $D_2^*N$ interactions can be described by exchanging pion mesons. In the following, we briefly introduce the effective Lagrangians involved in the whole calculation.

According to the heavy quark limit and chiral symmetry, the Lagrangian of the charmed mesons in the $T$ doublet interacting with the pseudoscalar meson was constructed in Ref. \cite{Ding:2008gr}, which has the form
\begin{eqnarray}\label{lagd1d2}
\mathcal{L}_{TT\mathbb{P}}&=& ik\langle T^{(Q)\mu}_b{\rlap\slash A}_{ba}\gamma_5\bar{T}^{(Q)}_{a\mu}\rangle,
\end{eqnarray}
where $\langle...\rangle$ means trace over the $2\times 2$ matrices. The axial current ${A}_{\mu}$ is defined by $A_{\mu} = \frac{1}{2}(\xi^{\dag}\partial_{\mu}\xi-\xi\partial_{\mu}\xi^{\dag})$
with $\xi=\exp(i\mathbb{P}/f_{\pi})$ and $f_{\pi}=132$ MeV, where
the matrix $\mathbb{P}$ reads as
\begin{eqnarray}
\mathbb{P} &=&\left(\begin{array}{cc}
\frac{\pi^0}{\sqrt{2}} &\pi^+  \nonumber\\
\pi^- &-\frac{\pi^0}{\sqrt{2}}
\end{array}\right).\nonumber
\end{eqnarray}
In addition, the multiplet field $T_a^{(Q)}$ is
\begin{eqnarray}
T_a^{(Q)\mu} &=& \frac{1+\rlap\slash v}{2}\left[P_{2a}^{*(Q)\mu\nu}\gamma_{\nu}
-\sqrt{\frac{3}{2}}P_{1a\nu}^{(Q)}\gamma_5(g^{\mu\nu}-\frac{1}{3}\gamma^{\nu}(\gamma^{\mu}-v^{\mu}))\right]\nonumber
\end{eqnarray}
with $v=(1,\vec{0})$, which is composed of axial vector $P_1$ and tensor $P_2$ with $P_1^T=(D_1^0,D_1^+)$ and $P_2^{*T}=(D_2^{*0},D_2^{*+})$. The conjugate field satisfies the relation $\bar{T}_a^{(Q)\mu}=\gamma_0T_a^{(Q)\mu\dag}\gamma_0$.
By expanding Eq. (\ref{lagd1d2}), we can further obtain the effective Lagrangian of $D_1(D_2^*)$ interacting with a pion, i.e.,
\begin{eqnarray}
 \mathcal{L}_{D_1D_1\mathbb{P}} &=&
            i\frac{5k}{3f_{\pi}}v^{\mu}\varepsilon_{\mu\nu\alpha\beta}
            D_{1b}^{\nu}D_{1a}^{\beta\dag}\partial^{\alpha}\mathbb{P}_{ba},\\
 \mathcal{L}_{D_1D_2^*\mathbb{P}} &=& -\sqrt{\frac{2}{3}}\frac{k}{f_{\pi}}
           \left(D_{2b}^{*\mu\nu}D_{1a\mu}^{\dag}+D_{1b\mu}D_{2a}^{*\mu\nu\dag}\right)
           \partial_{\nu}\mathbb{P}_{ba},\\
 \mathcal{L}_{D_2^*D_2^*\mathbb{P}} &=&   -2i\frac{k}{f_{\pi}}
            v^{\lambda}\varepsilon_{\lambda\nu\alpha\sigma}
            D_{2b}^{*\mu\nu}D_{2a\mu}^{*\sigma\dag}\partial^{\alpha}\mathbb{P}_{ba},
\end{eqnarray}
where the coupling constant $k$ is taken as the same value as that of $g$ as suggested in Refs. \cite{Falk:1992cx,Liu:2008xz}, which is the coupling constant of the charmed mesons in the $H=(0^-,1^-)$ doublet coupling with the pseudoscalar meson. However, different groups gave different results of $g$, i.e., $g=0.75$ estimated in the quark model \cite{Falk:1992cx}, $g=0.6$ listed in Ref. \cite{Bardeen:2003kt}, and $g=0.59\pm 0.07\pm 0.01$ extracted by the experimental width of $D^*$ \cite{Isola:2003fh}.
In this work, we take $g=0.59$ constrained by experiment to present the numerical result.

The effective Lagrangian of $NN\pi$ is
\begin{eqnarray}
\mathcal{L}_{NN\mathbb{P}} &=&
            -\frac{g_{\pi NN}}{\sqrt{2}m_{N}}\bar{N}_b\gamma^{\mu}\partial_{\mu}\mathbb{P}_{ba}
            \gamma_5N_{a}
\end{eqnarray}
with $N^T=(p,n)$  and the coupling constants $g^2_{\pi NN}/(4\pi)=13.6$ \cite{Machleidt:2000ge,Cao:2010km}.

In the above expressions, $P_1$, $P_2^*$ and $N$ satisfy the normalization relations $\langle 0|P_1^{\mu}|Q\bar{q}(1^+)\rangle=\epsilon^{\mu}\sqrt{M_{P_1}}$, $\langle 0|P_2^{*\mu\nu}|Q\bar{q}(2^+)\rangle=\zeta^{\mu\nu}\sqrt{M_{P_2^*}}$, and $\langle 0|N|qqq(1/2^+)\rangle=\sqrt{2M_{N}}\left(\left(1-\frac{\vec{p}^2}{8m_{N}^2}\right)\chi^{\dag},\frac{\vec{\sigma}\cdot\vec{p}}{2m_{N}}\chi^{\dag}\right)^T$, respectively.

\subsection{The OPE effective potential}

In the following, we deduce the OPE effective potential of these discussed molecular systems, where the Breit approximation is adopted. The involved scattering processes in the deduction include $D_1N\rightarrow D_1N$, $D_1N\rightarrow D_2^*N$, and $D_2^*N\rightarrow D_2^*N$. With the effective Lagrangians listed in Sec. \ref{subsec2}, we can write out the corresponding scattering amplitudes. And then, the effective potential in the momentum space can be related to the obtained scattering amplitude via the Breit approximation. Here, taking the $D_1N\rightarrow D_1N$ process as an example, we give its effective potential in the momentum space, i.e.,
\begin{eqnarray}\label{Breit}
\mathcal{V}_{E}^{D_1N\rightarrow D_1N}(\vec{q}) = -\frac{\mathcal{M}(D_1N\rightarrow D_1N)}
          {\sqrt{\prod_i2M_i\prod_f2M_f}},
\end{eqnarray}
where $\mathcal{M}(D_1N\rightarrow D_1N)$ denotes the amplitude of the $D_1N\rightarrow D_1N$ scattering process by exchanging pions. And $M_i$ and $M_f$ are the masses of the initial and final states, respectively.
By performing the Fourier transformation, we finally obtain the effective potential in the coordinate space:
\begin{eqnarray}\label{form fact}
\mathcal{V}_{E}^{D_1N\rightarrow D_1N}(\vec{r}) =
          \int\frac{d^3\vec{p}}{(2\pi)^3}e^{i\vec{p}\cdot\vec(r)}
          \mathcal{V}_{E}^{D_1N\rightarrow D_1N}(\vec{q})\mathcal{F}^2(q^2,m_E^2),
\end{eqnarray}
where $\mathcal{F}(q^2,m_E^2)$ is the monopole form factor with the form $\mathcal{F}(q^2,m_E^2)=(\Lambda^2-m_E^2)/(\Lambda^2-q^2)$. $m_E$ denotes the exchanged meson mass. Here, $\Lambda$ is the phenomenological parameter around 1 GeV.

First, we get the OPE scattering amplitudes corresponding to the $D_1N\rightarrow D_1N$, $D_1N\rightarrow D_2^*N$, and $D_2^*N\rightarrow D_2^*N$ processes, which are
\begin{eqnarray}
i\mathcal{M}_{D_1N\rightarrow D_1N} &=&
         i\left\{i\frac{5k}{3f_{\pi}}v^{\mu}\varepsilon_{\mu\nu\alpha\beta}
         m_{D_1}\epsilon_1^{\nu}\epsilon_3^{\beta\dag}(iq^{\alpha})\right\}\nonumber\\
         &&\times\frac{i}{q^2-m_{\pi}^2}
         i\left\{-\frac{g_{\mathbb{P}NN}}{\sqrt{2}m_{N}}\bar{u}\gamma^{\rho}(-iq_{\rho})\gamma_{5}u
         \right\}\nonumber\\
     &\simeq& -\left\{\frac{5k}{3f_{\pi}}m_{D_1}\frac{g_{\mathbb{P}NN}}{\sqrt{2}m_{N}}2m_N\right\}
     \frac{1}{\vec{q}^2+m_{\pi}^2}\nonumber\\
         &&\times\left[\left(\vec{\epsilon_1}^{\lambda_1}\times\vec{\epsilon_3}^{\lambda_3\dag}\right)\cdot \vec{q}\right]\left(\chi^{\dag}(\vec{\sigma}\cdot\vec{q})\chi\right),
\end{eqnarray}
\begin{eqnarray}
&&i\mathcal{M}_{D_1N\rightarrow D_2^*N}\nonumber\\&&=
           i\left\{-\sqrt{\frac{2}{3}}\frac{k}{f_{\pi}}\sqrt{m_{D_1}m_{D_2^*}}
           \epsilon_{1\mu}\zeta_3^{\mu\nu\dag}(iq_{\nu})\right\}\nonumber\\
         &&\quad\times\frac{i}{{q}^2-m_{\pi}^2}
           i\left\{-\frac{g_{\mathbb{P}NN}}{\sqrt{2}m_{N}}\bar{u}\gamma^{\rho}(-iq_{\rho})\gamma_{5}u
           \right\}\nonumber\\
       &&\simeq -i\left\{\sqrt{\frac{2}{3}}\frac{k}{f_{\pi}}\sqrt{m_{D_1}m_{D_2^*}}
           \frac{g_{\mathbb{P}NN}}{\sqrt{2}m_{N}}2m_{N}\right\}\frac{1}{\vec{q}^2+m_{\pi0}^2}\nonumber\\
           &&\quad\times\sum_{\lambda}\langle1,\lambda;1,\lambda_3-\lambda|2,\lambda_3\rangle\nonumber\\
           &&\quad\times\left[(\vec{\epsilon}_1^{\lambda_1}\cdot\vec{\epsilon}_3^{\lambda\dag})
           (\vec{\epsilon}_3^{(\lambda_3-\lambda)\dag}\cdot\vec{q})\right]
           \left(\chi^{\dag}(\vec{\sigma}\cdot\vec{q})\chi\right),
\end{eqnarray}
and
\begin{eqnarray}
&&i\mathcal{M}_{D_2^*N\leftrightarrow D_2^*N} \nonumber\\&&=
            i\left\{-2i\frac{k}{f_{\pi}}m_{D_2^*}v^{\lambda}\varepsilon_{\lambda\nu\alpha\sigma}
            \zeta_1^{\mu\nu}\zeta_{3\mu}^{\sigma\dag}(iq^{\alpha})\right\}\nonumber\\
         &&\quad\times\frac{i}{{q}^2-m_{\pi}^2}
            i\left\{-\frac{g_{\mathbb{P}NN}}{\sqrt{2}m_{N}}\bar{u}\gamma^{\rho}(-iq_{\rho})\gamma_{5}u
           \right\}\nonumber\\
         &&\simeq -\left\{2\frac{k}{f_{\pi}}m_{D_2^*}\frac{g_{\mathbb{P}NN}}{\sqrt{2}m_{N}}2m_N\right\}
         \frac{1}{\vec{q}^2+m_{\pi}^2}\nonumber\\
         &&\quad\times\sum_{\lambda, \lambda'}\langle1,\lambda;1,\lambda_1-\lambda|2,\lambda_1\rangle
         \langle1,\lambda';1,\lambda_3-\lambda'|2,\lambda_3\rangle\nonumber\\
           &&\quad\times(\vec{\epsilon}_1^{\lambda}\cdot\vec{\epsilon}_3^{\lambda'\dag})
           \left[\left(\vec{\epsilon}_1^{\lambda_1-\lambda}\times
           \vec{\epsilon}_3^{(\lambda_3-\lambda')\dag}\right)\cdot\vec{q}\right]
           \times\left(\chi^{\dag}(\vec{\sigma}\cdot\vec{q})\chi\right).
\end{eqnarray}

With the above preparation, we obtain the general expressions of the OPE potentials of these discussed $D_1N$ and $D_2^*N$ molecular systems with different $J$ quantum numbers
\begin{eqnarray}
\label{D10.5}
\mathcal{V}_{D_1N}^{J=1/2} &=&C_{{1}}
\left(\begin{array}{cc}
-2Z          &-\sqrt{2}T\\
-\sqrt{2}T   &Z-2T
\end{array}\right)\mathcal{G},\label{h1}\\
\mathcal{V}_{D_1N}^{J=3/2} &=& C_{{1}}
\left(\begin{array}{ccc}
Z   &T     &2T\\
T   &-2Z   &-T\\
2T  &-T    &Z
\end{array}\right)\mathcal{G},\label{D11.5}\\
\mathcal{V}_{D_2^*N}^{J=3/2} &=& C_{{2}}
\left(\begin{array}{ccc}
-\frac{3}{2}Z   &-\frac{3}{5}T     &-\frac{3\sqrt{21}}{10}T\\
-\frac{3}{5}T   &-\frac{3}{2}Z   &-\frac{3\sqrt{21}}{14}T\\
-\frac{3\sqrt{21}}{10}T  &-\frac{3\sqrt{21}}{14}T    &Z-\frac{4}{7}T
\end{array}\right)\mathcal{G},\label{D21.5}\nonumber\\
\\
\mathcal{V}_{D_2^*N}^{J=5/2} &=& C_{{2}}
\left(\begin{array}{ccc}
Z   &\frac{3\sqrt{14}}{10}T     &\frac{2\sqrt{14}}{5}T\\
\frac{3\sqrt{14}}{10}T   &-\frac{3}{2}Z-\frac{3}{7}T   &-\frac{3}{7}T\\
\frac{2\sqrt{14}}{5}T  &-\frac{3}{7}T    &Z+\frac{4}{7}T
\end{array}\right)\mathcal{G},\label{D22.5}\nonumber\\\label{h4}
\end{eqnarray}
where the S-D mixing effect is considered, which is the reason why the obtained OPE effective potentials for the $D_1N$ and $D_2^*N$ systems have the matrix form. To distinguish the effective potentials
listed in Eqs. (\ref{h1})-(\ref{h4}), we use the subscripts $D_1N$ and $D_2^*N$ and the superscripts $J=1/2, \,3/2,\,5/2$. In addition,
we need to specify that the deduced OPE effective potentials in Eqs. (\ref{h1})-(\ref{h4}) do not include the coupled channel effect.

If considering the coupled channel effect in the deduction of the effective potential, we notice that there exists the coupling of the $D_1N$ and $D_2^*N$ channels $J=1/2$ or $J=3/2$. Thus, we give the OPE effective potential of the $D_1(D_2^*)N$ systems with $J= 1/2$ and $J= 3/2$, i.e.,
\begin{widetext}
\begin{eqnarray}
\mathcal{V}_{}^{J=1/2} &=& \left(\begin{array}{cccc}
-2C_{{1}}Z          &-\sqrt{2}C_{{1}}T    &\frac{1}{\sqrt{5}}C_{{3}}T_0     &6\sqrt{\frac{2}{15}}C_{{3}}T_0\\
-\sqrt{2}C_{{1}}T   &C_{{1}}(Z-2T)          &C_{{3}}(\frac{\sqrt{10}}{6}Z_0+\sqrt{\frac{2}{5}}T_0)     &-\frac{3}{\sqrt{15}}C_{{3}}T_0\\
\frac{1}{\sqrt{5}}C_{{3}}T_0             &C_{{3}}(\frac{\sqrt{10}}{6}Z_0+\sqrt{\frac{2}{5}}T_0)   &C_{{2}}(-\frac{3}{2}Z+\frac{3}{5}T)      &-\frac{3\sqrt{6}}{10}C_{{2}}T\\
6\sqrt{\frac{2}{15}}C_{{3}}T_0          &-\frac{3}{\sqrt{15}}C_{{3}}T_0    &-\frac{3\sqrt{6}}{10}C_{{2}}T              &C_{{2}}(Z-\frac{8}{5}T)
\end{array}\right)\mathcal{G},\label{D1D20.5}
\end{eqnarray}

\begin{small}
\begin{eqnarray}
\mathcal{V}_{}^{J=3/2} &=&\mathcal{G}\nonumber\\
&&\times\left(\begin{array}{cccccc}
C_{{1}}Z  &C_{{1}}T   &2C_{{1}}T   &\frac{1}{\sqrt{10}}C_{{3}}Z_0    &-\sqrt{\frac{2}{5}}C_{{3}}T_0  &-\sqrt{\frac{21}{10}}C_{{3}}T_0\\
C_{{1}}T  &-2C_{{1}}Z  &-C_{{1}}T  &-\frac{1}{\sqrt{10}}C_{{3}}T_0   &C_{{3}}(-\frac{\sqrt{10}}{75}Z_0+\frac{1}{\sqrt{10}}T_0)  &C_{{3}}(\frac{2\sqrt{210}}{75}Z_0-\sqrt{\frac{24}{35}}T_0)\\
2C_{{1}}T &-C_{{1}}T  &C_{{1}}Z
 &-\sqrt{\frac{2}{5}}C_{{3}}T_0  &\frac{31}{15 \sqrt{10}}C_{{3}}Z_0   &C_{{3}}(-\frac{\sqrt{210}}{75}Z_0-\sqrt{\frac{15}{14}}T_0)\\
\frac{1}{\sqrt{10}}C_{{3}}Z_0    &-\frac{1}{\sqrt{10}}C_{{3}}T_0  &-\sqrt{\frac{2}{5}}C_{{3}}T_0
&-\frac{3}{2}C_{{2}}Z   &-\frac{3}{5}C_{{2}}T     &-\frac{3\sqrt{21}}{10}C_{{2}}T \\
-\sqrt{\frac{2}{5}}C_{{3}}T_0   &C_{{3}}(-\frac{\sqrt{10}}{75}Z_0+\frac{1}{\sqrt{10}}T_0)  &\frac{31}{15 \sqrt{10}}C_{{3}}Z_0
&-\frac{3}{5}C_{{2}}T   &-\frac{3}{2}C_{{2}}Z   &-\frac{3\sqrt{21}}{14}C_{{2}}T\\
-\sqrt{\frac{21}{10}}C_{{3}}T_0  &C_{{3}}(\frac{2\sqrt{210}}{75}Z_0-\sqrt{\frac{24}{35}}T_0)   &C_{{3}}(-\frac{\sqrt{210}}{75}Z_0-\sqrt{\frac{15}{14}}T_0)   &-\frac{3\sqrt{21}}{10}C_{{2}}T  &-\frac{3\sqrt{21}}{14}C_{{2}}T    &C_{{2}}(Z-\frac{4}{7}T)
\end{array}\right).\label{D1D21.5}\nonumber\\
\end{eqnarray}
\end{small}
\end{widetext}

In Eqs. (\ref{h1})-(\ref{D1D21.5}), the isospin factor $\mathcal{G}$ is taken as 3/2 for the isoscalar state, and -1/2 for the isovector state. Additionally, we also define
 $C_{{1}}= -\frac{5}{18\sqrt{2}}\frac{k g_{\pi NN}}{f_{\pi}m_{N}}$,
$C_{{2}}= -\frac{1}{3\sqrt{2}}\frac{k g_{\pi NN}}{f_{\pi}m_{N}}$, and
$C_{{3}} =  \frac{\sqrt{3}}{18}\frac{k g_{\pi NN}}{f_{\pi}m_{N}}$.
The $Z$ and $T$ in Eqs. (\ref{h1})-(\ref{D1D21.5}) are the abbreviation of functions $Z(\Lambda,m_{\pi},\vec{r})$ and $T(\Lambda,m_{\pi},\vec{r})$, which have the forms
\begin{eqnarray}
Y(\Lambda,m_{\rho},{r}) &=&\frac{1}{4\pi r}(e^{-mr}-e^{-\Lambda r})-\frac{\Lambda^2-m^2}{8\pi \Lambda}e^{-\Lambda r},\label{yy}\\
Z(\Lambda,m,r) &=& \nabla^2Y(\Lambda,m,r)=\frac{1}{r^2}\frac{\partial}{\partial r}r^2\frac{\partial}{\partial r}Y(\Lambda,m,r),\label{zz}\\
T(\Lambda,m,r) &=& r\frac{\partial}{\partial r}\frac{1}{r}\frac{\partial}{\partial r}Y(\Lambda,m,r).\label{tt}
\end{eqnarray}
And we also have functions $Z_0=Z(\Lambda_{0},m_{\pi{0}},\vec{r})$ and $T_0=T(\Lambda_{0},m_{\pi{0}},\vec{r})$ with
\begin{eqnarray}
\Lambda_0^2=\Lambda^2-\left(\frac{m_{D_2^*}^2-m_{D_1}^2}{2(m_N+m_{D_2}^*)}\right)^2,\quad
m_{\pi{0}}^2=m_{\pi}^2-\left(\frac{m_{D_2^*}^2-m_{D_1}^2}{2(m_N+m_{D_2}^*)}\right)^2.\nonumber
\end{eqnarray}

With these OPE effective potentials listed in Eqs. (\ref{h1})-(\ref{D1D21.5}), we can find the bound state solutions for these $D_1N$ and $D_2^*N$ molecular states by solving the coupled channel Schr$\ddot{\text{o}}$dinger equation. Here, the kinetic terms are
\begin{eqnarray}
K_{D_1N}^{J=1/2} &=& \text{diag}\left(-\frac{1}{2m_1}\nabla^2_0,-\frac{1}{2m_1}\nabla^2_1\right),\nonumber\\
K_{D_1N}^{J=3/2} &=& \text{diag}\left(-\frac{1}{2m_1}\nabla^2_0,
                   -\frac{1}{2m_1}\nabla^2_1,-\frac{1}{2m_1}\nabla^2_1\right),\nonumber\\
K_{D_2^*N}^{J=3/2} &=& \text{diag}\left(-\frac{1}{2m_2}\nabla^2_0,
                   -\frac{1}{2m_2}\nabla^2_1,-\frac{1}{2m_2}\nabla^2_1\right),\nonumber\\
K_{D_2^*N}^{J=5/2} &=& \text{diag}\left(-\frac{1}{2m_2}\nabla^2_0,
                   -\frac{1}{2m_2}\nabla^2_1,-\frac{1}{2m_2}\nabla^2_1\right),\nonumber
\end{eqnarray}
which correspond to the effective potentials in Eqs. (\ref{h1})-(\ref{h4}), respectively.

The kinetic terms corresponding to Eqs. (\ref{D1D20.5}) and (\ref{D1D21.5}) are
\begin{eqnarray}
&&K_{}^{J=1/2}\nonumber\\ &&= \text{diag}\bigg(
       -\frac{1}{2m_1}\nabla^2_0,-\frac{1}{2m_1}\nabla^2_1,\nonumber\\
                 &&\quad-\frac{1}{2m_2}\nabla^2_1+\Delta{m},
                 -\frac{1}{2m_2}\nabla^2_1+\Delta{m}\bigg),\label{kinetic1}\\
&&K_{}^{J=3/2} \nonumber\\&&= \text{diag}\bigg(-\frac{1}{2m_1}\nabla^2_0,
                   -\frac{1}{2m_1}\nabla^2_1,
                   -\frac{1}{2m_1}\nabla^2_1,\nonumber\\
                   &&\quad-\frac{1}{2m_2}\nabla^2_0+\Delta{m},
                   -\frac{1}{2m_2}\nabla^2_1+\Delta{m}
                   ,-\frac{1}{2m_2}\nabla^2_1+\Delta{m}\bigg),\nonumber\\\label{kinetic2}
\end{eqnarray}
respectively, where $\nabla_0^2=\frac{1}{r^2}\frac{\partial}{\partial{r}}r^2\frac{\partial}{\partial{r}}$,
         $\nabla_1^2=\nabla_0^2-6/{r^2}$, $m_1=\frac{m_{D_1}*m_{N}}{m_{D_1}+m_{N}}$, $m_2=\frac{m_{D_2^*}*m_{N}}{m_{D_2^*}+m_{N}}$, and $\Delta{m}=m_{D_2^*}-m_{D_1}$.

\renewcommand{\arraystretch}{1.3}
\begin{table}[hbt]
\caption{The masses and the quantum numbers of the particles in these discussed systems. Here, these values are taken from the Particle Data Group \cite{Beringer:1900zz}.\label{value}}
\begin{tabular}{ccc}
\midrule[1pt]
~~~Hadrons~~~          &~~~$I(J^P)$~~~          &~~~Mass (MeV)~~~\\\midrule[1pt]
 $D_1^{\pm}$     &$\frac{1}{2}(1^+)$     &2423.4\\
 $D_1^{0}$       &$\frac{1}{2}(1^+)$     &2421.3\\
 $D_2^{*\pm}$    &$\frac{1}{2}(2^+)$     &2464.4\\
 $D_2^{*0}$      &$\frac{1}{2}(2^+)$     &2462.6\\
 $p$           &$\frac{1}{2}(\frac{1}{2}^+)$     &938.27\\
 $n$           &$\frac{1}{2}(\frac{1}{2}^+)$     &939.56\\
 $\pi^{\pm}$         &$1(0^-)$                      &139.57\\
 $\pi^{0}$         &$1(0^-)$                      &134.98\\\midrule[1pt]
\end{tabular}
\end{table}

\begin{figure*}[htbp]
\begin{tabular}{ccc}
\multicolumn{2}{c}{{\Large Case I}}&\\
\includegraphics[scale=0.42]{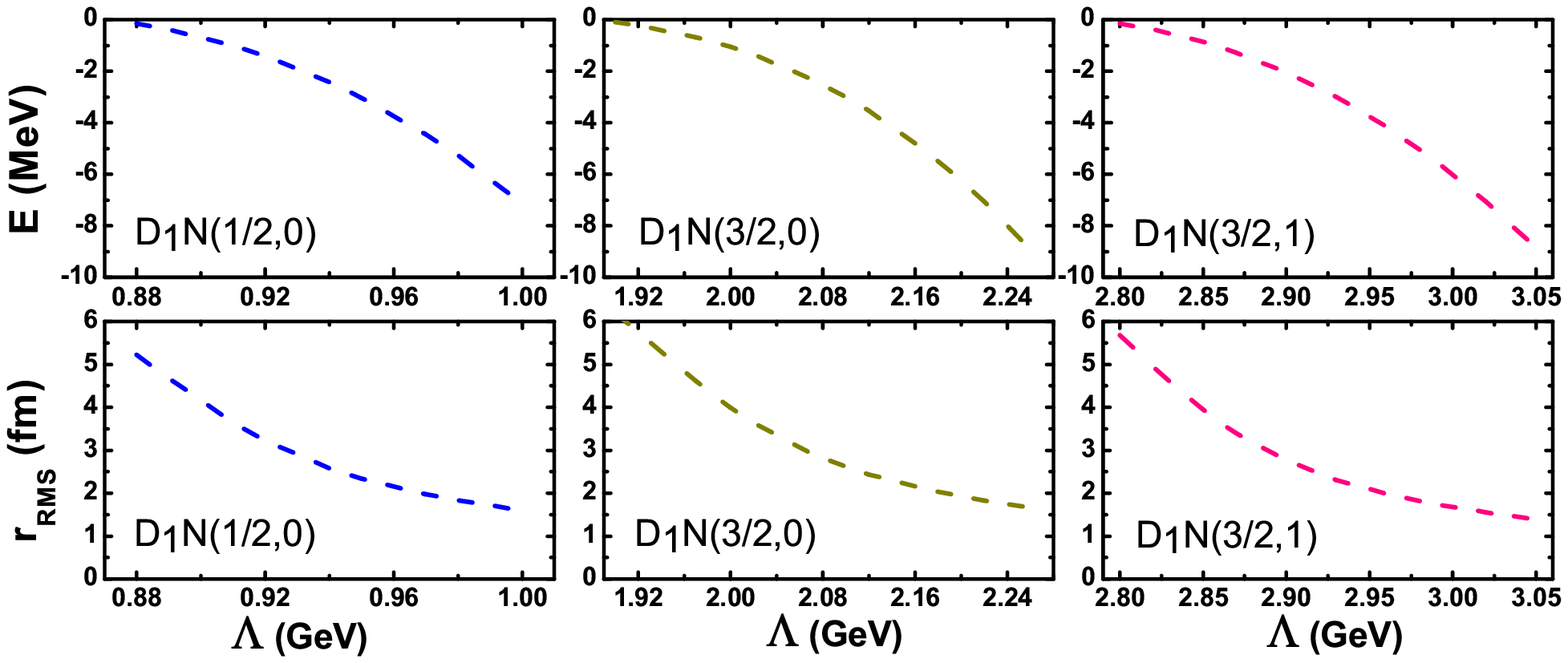}&\includegraphics[scale=0.42]{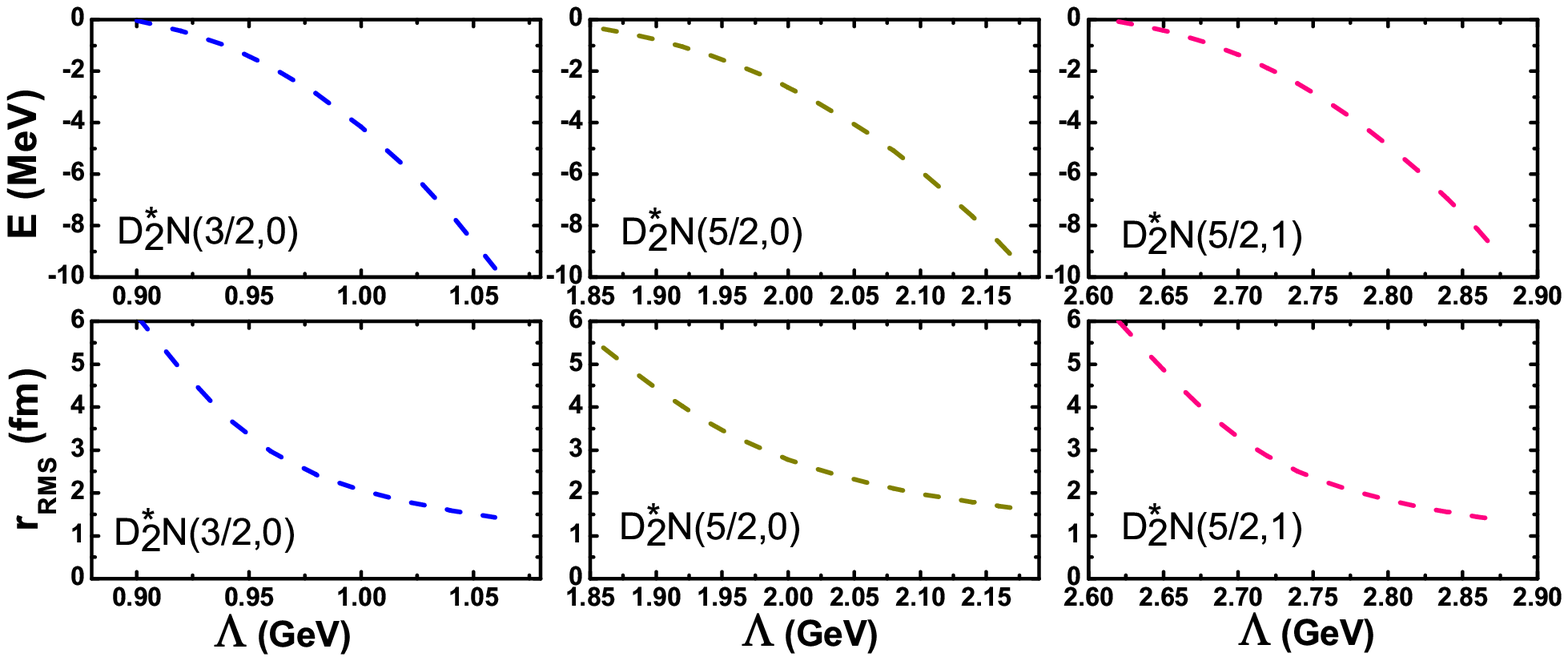}\\
\multicolumn{2}{c}{{\Large Case II}}&\\
\includegraphics[scale=0.42]{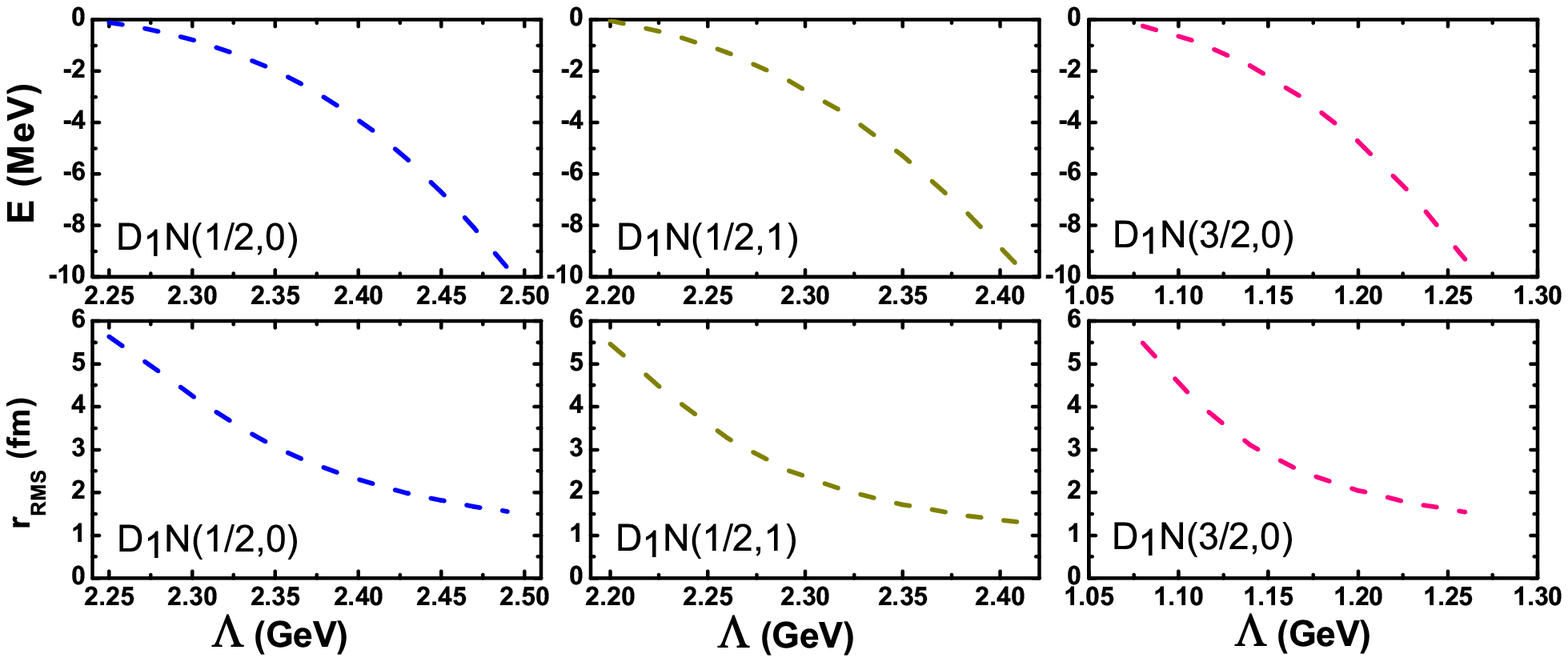}&\includegraphics[scale=0.42]{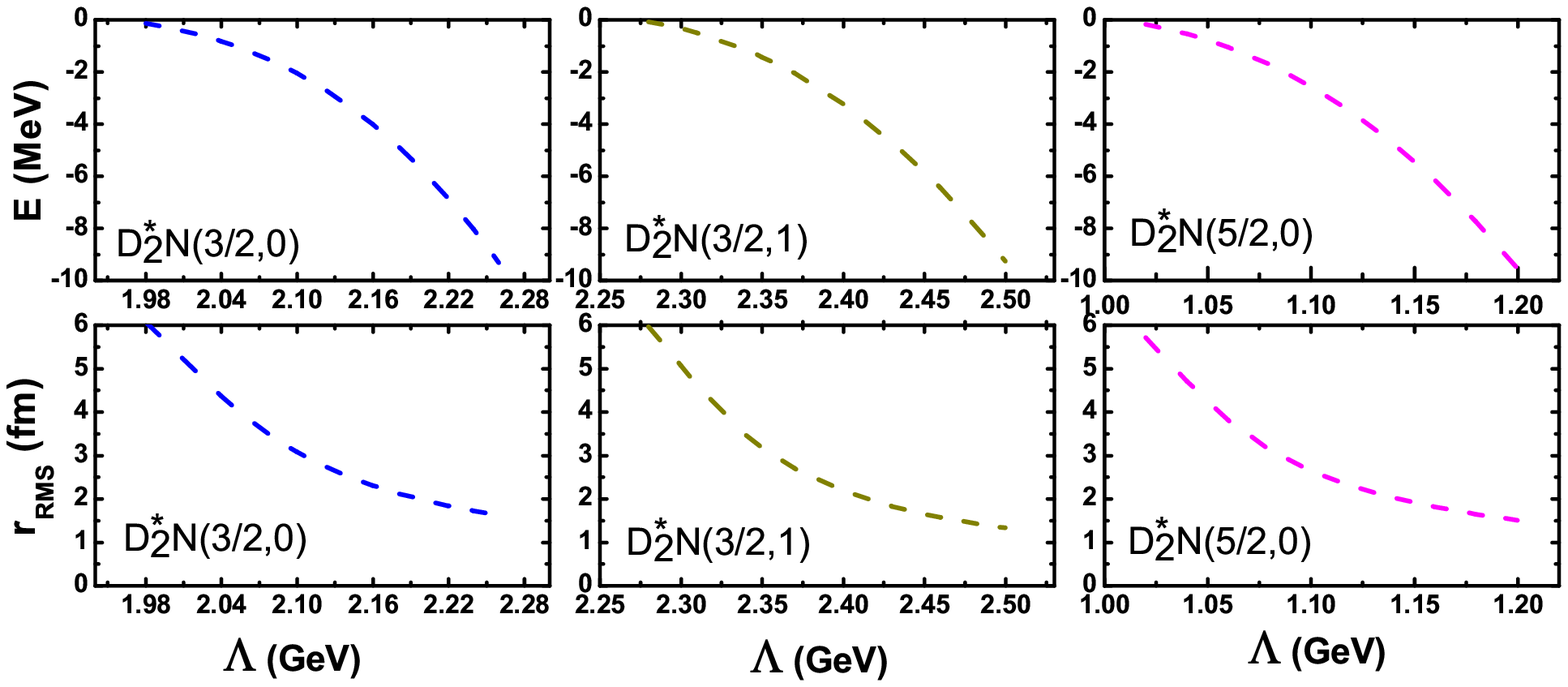}
\end{tabular}
\caption{The $\Lambda$ dependence of the obtained bound state solutions (the binding energy $E$ and the root-mean-square radius $r_{RMS}$) for eight $D_1N$ and $D_2^*N$ systems with typical coupling constants $k=0.59$, $g^2_{\pi NN}/(4\pi)=13.6$. We discuss two cases of the signs of $g_{\pi NN}\,k$ as introduced in this work. $E$, $r_{RMS}$, and $\Lambda$ are in units of MeV, fm, and GeV, respectively.\label{sp}}
\end{figure*}

\begin{table}[hbtp]
\caption{The typical values of the obtained bound state solutions (the binding energy $E$ and the root-mean-square radius $r_{RMS}$) for eight $D_1N$ and $D_2^*N$ systems corresponding to Fig. \ref{sp}. Here, for showing the dependence of numerical results on $\Lambda$, we list three typical results for each of the systems corresponding to three typical $\Lambda$ values. }\label{D1D1}
\begin{tabular}{c|c|ccc|ccc}
\midrule[1pt]
&&\multicolumn{3}{c}{Case I}&\multicolumn{3}{c}{Case II}\\
\midrule[1pt]
  States     &$(J,I)$
 &$\Lambda$   &$E$ &$r_{RMS}$&$\Lambda$   &$E$  &$r_{RMS}$\\ \hline
 \multirow{12}*{$D_1N$ }   &\multirow{3}*{$(1/2,0)$}
     &0.88      &-0.15   &5.22       &2.30  &-0.78   &4.26   \\
 &   &0.94      &-2.42   &2.58     &2.40  &-3.90   &2.29 \\
 &   &1.00      &-7.15   &1.60     &2.50  &-10.54   &1.50  \\
 \cline{2-8}
 &\multirow{3}*{$(1/2,1)$}
     &--      &--   &--  &2.25  &-1.00   &3.58    \\
     &&--&--&--   &2.30  &-2.72   &2.35      \\
 &&--&--  &-- &2.35  &-5.30   &1.72      \\
\cline{2-8}
&\multirow{3}*{$(3/2,0)$}
     &2.00  &-1.06   &3.99   &1.10  &-0.63   &4.53    \\
 &   &2.10  &-3.00   &2.61  &1.15  &-2.18   &2.85    \\
 &   &2.20  &-6.24   &1.92    &1.20  &-4.75   &2.04  \\
 \cline{2-8}
  &\multirow{3}*{$(3/2,1)$}
     &2.85  &-0.86   &3.94   &--  &--   &--    \\
 &   &2.95  &-3.74   &2.08  &--  &--   &--    \\
 &   &3.05  &-8.86   &1.40   &-- &--   &--   \\
\midrule[1pt]
\multirow{12}*{$D_2^*N$ }   &\multirow{3}*{$(3/2,0)$}
     &0.92     &-0.43   &4.86       &2.00  &-0.32   &5.50   \\
 &   &0.96     &-1.82   &2.96     &2.10  &-2.05   &3.08 \\
 &   &1.00      &-4.17   &2.06     &2.20  &-5.81   &1.97  \\
 \cline{2-8}
 &\multirow{3}*{$(3/2,1)$}
     &--      &--   &--  &2.30  &-0.33   &5.06    \\
     &&--&--&--   &2.40  &-3.24   &2.18      \\
 &&--&--  &-- &2.50  &-9.28   &1.33      \\
\cline{2-8}
&\multirow{3}*{$(5/2,0)$}
     &1.90  &-0.78   &4.44   &1.05  &-0.77   &4.23    \\
 &   &2.00  &-2.63   &2.78  &1.10  &-2.54   &2.64    \\
 &   &2.10  &-5.89   &1.98    &1.15  &-5.43   &1.91  \\
 \cline{2-8}
  &\multirow{3}*{$(5/2,1)$}
     &2.65  &-0.42   &4.86   &--  &--   &--    \\
 &   &2.75  &-2.83   &2.36  &--  &--   &--    \\
 &   &2.85  &-7.55   &1.50   &-- &--   &--   \\
\bottomrule[1pt]
\end{tabular}
\end{table}

\section{numerical results}\label{sec3}

In Sec. \ref{sec2}, we have deduced the concrete expressions of the OPE effective potentials of the systems composed of a $D$ meson in the $T$ doublet and a nucleon. The corresponding masses of the particles involved in our calculation are collected in Table \ref{value}. For numerically solving the coupled channel Schr$\ddot{\text{o}}$dinger equation and finding the bound state solution, we adopt the FESSDE program \cite{Abrashkevichn:1995cj,Abrashkevichn:1998cj}, where the binding energy and the corresponding root-mean-square radius will be given, which are dependent on the cutoff values in the monopole form factor.

Since the signs of $k$ and $g_{\pi NN}$ cannot be constrained, we discuss two cases, i.e., the sign of $g_{\pi NN}\,k$ is either positive or negative, which correspond to case I and case II, respectively.

\subsection{Pure $D_1N$ and $D_2^*N$ systems without considering coupled-channel effect}

We present the $\Lambda$ dependence of the obtained binding energy and the corresponding root-mean-square radius
of the $D_1N$ states with $J=1/2,\,3/2$ and isospin $I=0,\,1$ and the $D_2^*N$ states with $J=3/2,\,5/2$ and isospin $I=0,\,1$ 
in Fig. \ref{sp}, where the corresponding typical values are collected in Table \ref{D1D1} for the convenience of the  reader. 

\subsubsection{Case I}\label{22}

Under case I, we cannot find the bound state solutions for the $D_1N$ state with $J=1/2$ and $I=1$, and the $D_2^*N$ state with $J=3/2$ and $I=1$, which means that these two systems cannot be bound as the molecular state.
For the remaining six systems with the $(J,I)$ quantum numbers listed in the second column of Table \ref{D1D1}, we can find their bound state solutions with the corresponding cutoff $\Lambda$ (see Fig. 1 for more details of the binding energy dependent on $\Lambda$).

Although there exist the systems with the bound state solution, we further select more possible candidates as the loosely molecular states among the above states with the bound state solution. According to the experience of studying the deuteron via the one boson exchange model \cite{Tornqvist:1993vu,Tornqvist:1993ng}, usually $\Lambda$ should be around 1 GeV. If taking this criterion, we find that the isoscalar $D_1N$ states with $J=1/2$ and the isoscalar $D_2^*N$ states with $J=3/2$ are the better candidates of the loosely molecular state than the other four systems.

Experimental search for these predicted molecular states is an interesting research topic. Thus, in the following, we discuss their decay behavior.  For the isoscalar $D_1N$ molecular state with $J=1/2$ which is named as $D_1N[0(\frac{1}{2}^+)]$, its allowed decay mods include $DN$, $D^*N$, $\Lambda_c\eta$, $\Lambda_c\omega$, $\Sigma_c\pi$, $\Sigma_c\rho$ and $\Lambda_c\sigma$. Since $\rho$ and $\sigma$ can dominantly decay into $2\pi$, $D_1N[0(\frac{1}{2}^+)]$ decays into $\Sigma_c\pi\pi$ and $\Lambda_c\pi\pi$ are two important three-body decay channels. 

For the isoscalar $D_2^*N$ molecular state with $J=3/2$ (named as $D_2^*N[0(\frac{3}{2}^+)]$), $DN$, $D^*N$, $\Lambda_c\eta$, $\Lambda_c\omega$, $\Sigma_c\pi$, and $\Sigma_c\rho$ are its main two-body decay channels, where the three-body decay $\Sigma_c\pi\pi$ induced by $\Sigma_c\rho$ mode is also allowed. 

\begin{figure*}[htbp]
\begin{tabular}{c}
{\Large Case I}\\
\includegraphics[scale=0.8]{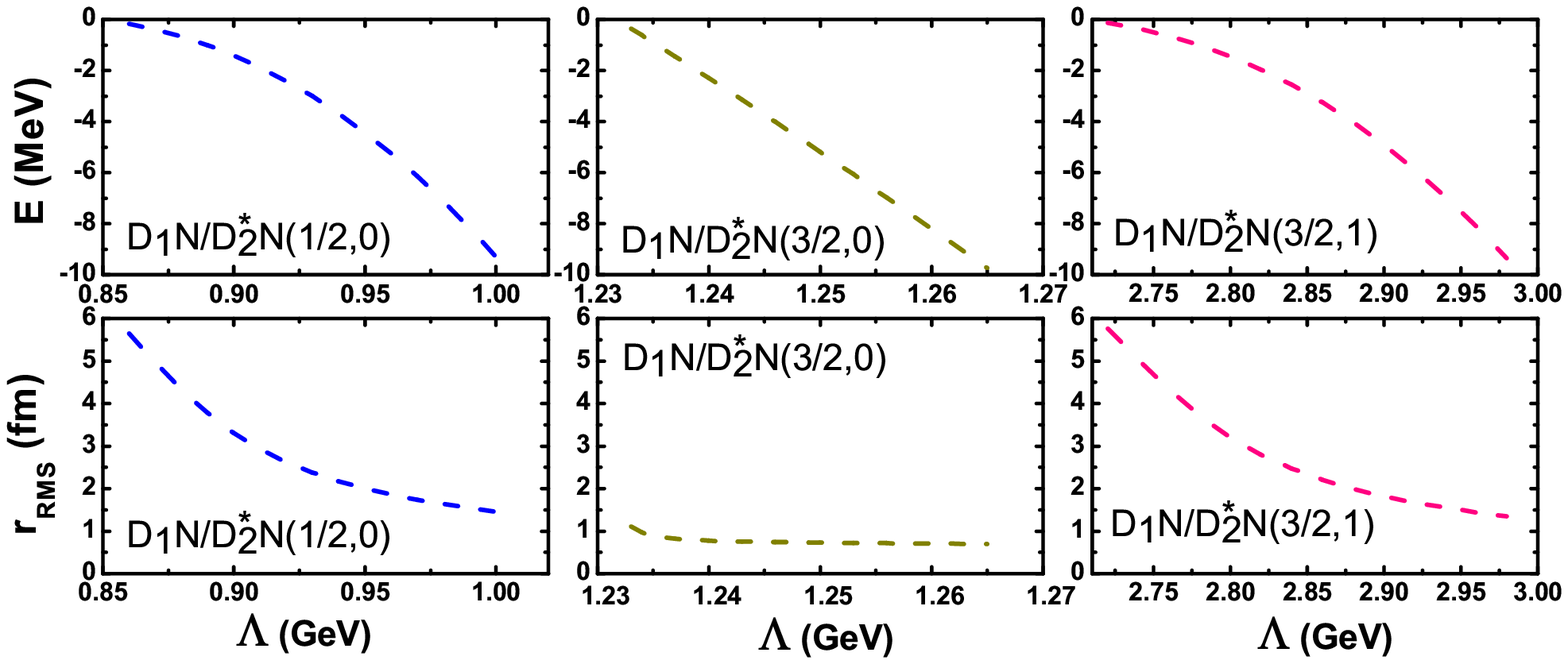}\\\\
{\Large Case II}\\
\includegraphics[bb=250 10 330 280,scale=0.8]{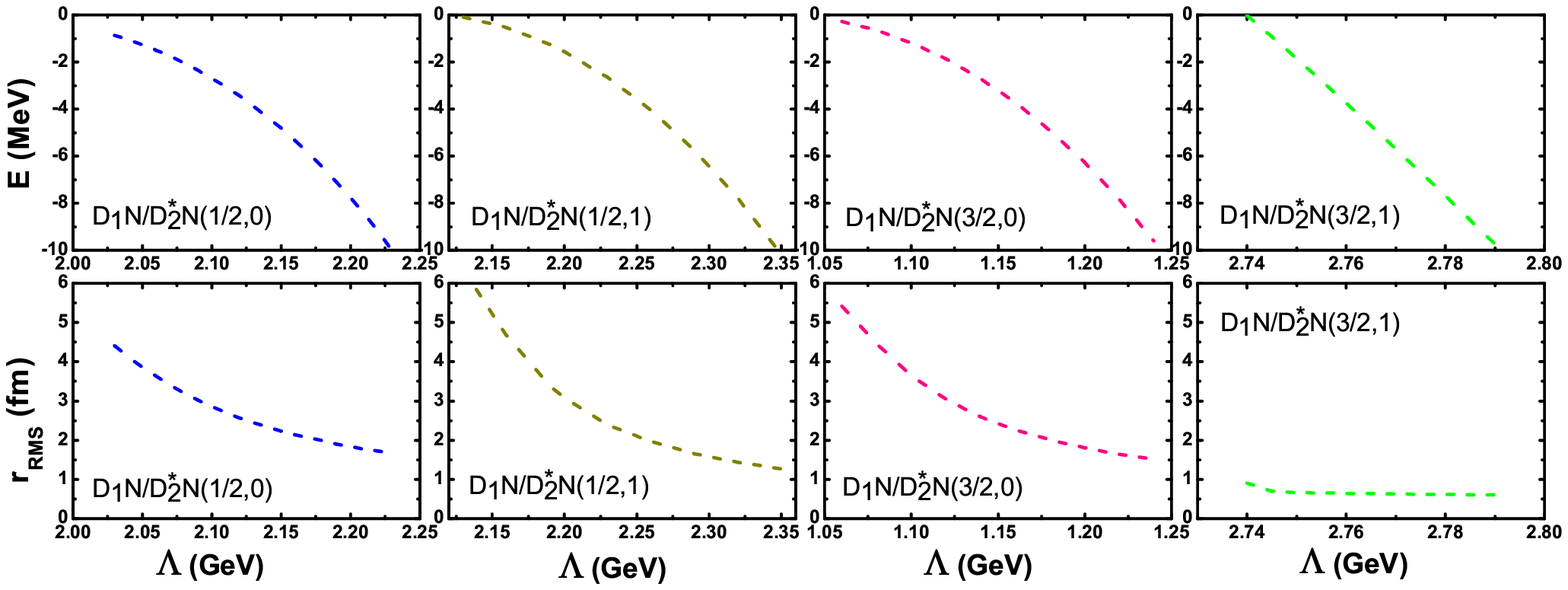}
\end{tabular}
\caption{The $\Lambda$ dependence of the obtained bound state solutions (the binding energy $E$ and the root-mean-square radius $r_{RMS}$) for the $D_1N/D_2^*N$ systems with typical coupling constants $k=0.59$, $g^2_{\pi NN}/(4\pi)=13.6$. We discuss two cases of the signs of $g_{\pi NN}\,k$ as introduced in this work. $E$, $r_{RMS}$, and $\Lambda$ are in units of MeV, fm, and GeV, respectively.\label{sp1}}
\end{figure*}

\begin{table*}[hbtp]
\caption{The typical values of the obtained binding energy $E$ and the root-mean-square radius $r_{RMS}$ for the $D_1N/D_2^*N$ systems corresponding to Fig. \ref{sp1}. Here, $p_{i}$ denotes the probability of the different channels.}\label{D1D2}
\scriptsize
\begin{tabular}{c|c|ccc|cccccc|ccc|cccccc}
\midrule[1pt]
&&\multicolumn{9}{c}{Case I}&\multicolumn{9}{|c}{Case II}\\
\midrule[1pt]
  States     &$(J,I)$
 &$\Lambda$   &$E$ &$r_{RMS}$   &$p_1(\%)$  &$p_2(\%)$  &$p_3(\%)$
 &$p_4(\%)$  &$p_5(\%)$  &$p_6(\%)$
 &$\Lambda$ &$E$  &$r_{RMS}$   &$p_1(\%)$  &$p_2(\%)$  &$p_3(\%)$
 &$p_4(\%)$  &$p_5(\%)$  &$p_6(\%)$\\ \hline
 \multirow{12}*{$D_1N/D_2^*N$ }   &\multirow{3}*{$(1/2,0)$}
     &0.87      &-0.39   &4.96
     &99.35    &0.05    &0.00     &0.14    &-        &-
      &2.03      &-0.86   &4.41
     &94.59    &4.07    &0.13     &1.21    &-        &-\\
 &   &0.93      &-3.02   &2.38
     &98.85    &0.84    &0.00     &0.29    &-        &-
 &2.09      &-2.35   &3.02
     &91.27    &6.46    &0.22     &2.06    &-        &-\\
 &   &0.99      &-8.17   &1.54
     &98.58    &1.00    &0.01     &0.40    &-        &-
 &2.15      &-4.82   &2.24
     &87.80    &8.87    &0.32     &3.00    &-        &-\\
 \cline{2-20}
 &\multirow{3}*{$(1/2,1)$}
 &-- &--&--
 &-  &-  &-  &-  &-  &-
 &2.16      &-0.54   &4.67
     &99.58    &0.29    &0.00     &0.12    &-        &-\\
 &&--&--&--
 &-  &-  &-  &-  &-  &-
 &2.22      &-2.25   &2.58
     &99.30    &0.48    &0.005     &0.18    &-        &-\\
 &&--&--&--
 &-  &-  &-  &-  &-  &-
 &2.29      &-5.76   &1.65
     &99.06    &0.64    &0.007     &0.29    &-        &-\\
\cline{2-20}
&\multirow{3}*{$(3/2,0)$}
&1.235      &-0.91   &0.89
     &1.47      &0.03    &0.19     &96.26   &0.31    &1.72
&1.06      &-0.27    &5.41
     &97.64    &0.52     &1.70     &0.02    &0.03    &0.09\\
 & &1.245      &-3.73   &0.75
     &0.59      &0.02    &0.16     &97.14   &0.32    &1.77
 &1.12      &-1.87    &3.04
     &95.66    &0.95     &3.09     &0.05    &0.06    &0.19\\
 &&1.305      &-23.23   &0.63
     &0.23      &0.02    &0.11     &97.48   &0.34    &1.82
 &1.18      &-4.91    &2.00
     &94.11    &1.27     &4.15     &0.09    &0.09    &0.29\\
 \cline{2-20}
  &\multirow{3}*{$(3/2,1)$}
     &2.70      &-0.17   &5.63
     &98.93      &0.20    &0.74     &0.03   &0.02    &0.06
     &2.75      &-1.85    &0.67
     &0.31     &0.01     &0.08     &97.85    &0.25    &1.49\\
 &   &2.80      &-2.14   &2.69
     &97.50      &0.46    &1.73     &0.09   &0.05    &0.16
 &2.76      &3.75     &0.64
     &0.22     &0.01     &0.08     &97.92    &0.26    &1.51\\
 &   &2.90      &-6.30   &1.62
     &96.25      &0.68    &2.57     &0.16   &0.08    &0.26
 &2.82 &-16.19   &0.57
 &0.13  &0.01  &0.08  &97.94  &0.27  &1.57\\
\bottomrule[1pt]
\end{tabular}
\end{table*}

\subsubsection{Case II}\label{11}

In the following, we perform the discussion under case II. Our result shows that there do not exist bound state solutions for the $D_1N$ state with $J=3/2$ and $I=1$ and the $D_2^*N$ state with $J=5/2$ and $I=1$. For the remaining six $D_1N$ and $D_2^*N$ systems, we find their bound state solutions. If taking the same criteria of $\Lambda$ as that presented above, we can find two favorable molecular states, i.e., $D_1N$ with $J=3/2$ and $I=0$, and $D_2^*N$ with $J=5/2$ and $I=0$, which are named as $D_1N[0(\frac{3}{2}^+)]$ and $D_2^*N[0(\frac{5}{2}^+)]$, respectively. Here, the remaining four systems have the bound state solutions with the cutoff deviated from usual requirement $\Lambda\sim 1$ GeV. 

These results show that it is easy to form isoscalar $D_1N$ and $D_2^*N$ molecular
states compared with these isovector $D_1N/D_2^*N$
systems, which also hold for the case I.  
Thus, we suggest future experiments first carry out the search for isoscalar $D_1N$ and $D_2^*N$
molecular states.

The allowed two-body decays of $D_1N[0(\frac{3}{2}^+)]$ molecular state are $DN$, $D^*N$, $\Lambda_c\eta$, $\Lambda_c\omega$, $\Sigma_c\pi$, and $\Sigma_c\rho$, while $D_2^*N[0(\frac{5}{2}^+)]$ can decay into $D^*N$, $\Lambda_c\omega$, and $\Sigma_c\rho$. In addition, for both $D_1N[0(\frac{3}{2}^+)]$ and $D_2^*N[0(\frac{5}{2}^+)]$, there exists a three-body decay channel $\Sigma_c\pi\pi$.

\begin{table*}[htbp]
\caption{The allowed decay channels of the discussed $D_1N/D_2^*N$ molecular states with different quantum numbers. Here, we use $\surd$ to mark these allowed decays. \label{tabIV}}
\begin{tabular}{c|ccccccccccc}
\toprule[1pt]
		\multirow{2}{*}{Decay channels}&$D_1N[0(\frac{1}{2})^+]$&$D_1N[1(\frac{1}{2})^+]$&$D_2^*N[0(\frac{3}{2})^+]$&$D_2^*N[1(\frac{3}{2})^+]$&\multirow{2}{*}{$D_2^*N[0(\frac{5}{2})^+]$}&\multirow{2}{*}{$D_2^*N[1(\frac{5}{2})^+]$}\\

&$D_1N/D_2^*N[0(\frac{1}{2})^+]$&$D_1N/D_2^*N[1(\frac{1}{2})^+]$&$D_1N/D_2^*N[0(\frac{3}{2})^+]$&$D_1N/D_2^*N[1(\frac{3}{2})^+]$&&\\

\toprule[1pt]
$DN$             &$\surd$&$\surd$&$\surd$&$\surd$&&\\
$D^*N$           &$\surd$&$\surd$&$\surd$&$\surd$&$\surd$&$\surd$\\
$\Lambda_c\pi$   &       &$\surd$&       &$\surd$&&\\
$\Lambda_c\eta$  &$\surd$&       &$\surd$&       &&\\
$\Lambda_c\rho$  &       &$\surd$&       &$\surd$&&$\surd$\\
$\Lambda_c\omega$&$\surd$&       &$\surd$&       &$\surd$&\\
$\Sigma_c\pi$    &$\surd$&$\surd$&$\surd$&$\surd$&&\\
$\Sigma_c\eta$   &       &$\surd$&       &$\surd$&&\\
$\Sigma_c\rho$   &$\surd$&$\surd$&$\surd$&$\surd$&$\surd$&$\surd$\\
$\Sigma_c\omega$ &       &$\surd$&       &$\surd$&&$\surd$\\
$\Lambda_c\sigma$&$\surd$&&&&&\\\bottomrule[1pt]
\end{tabular}
\end{table*}

\subsection{Considering coupled-channel effect with $D_1N/D_2^*N$ systems}

In the following, we discuss the $D_1N/D_2^*N$ states with $J=1/2,\,3/2$ and $I=0,1$ by considering the coupled channel effect, where the numerical results are given in Table \ref{D1D2} under case I and case II. The $\Lambda$ dependence of the bound state solutions is shown in Fig. \ref{sp1} and Table \ref{tabIV}. 

\subsubsection{Case I}

For the $D_1N/D_2^*N$ systems with $J=1/2$ and $I=0,1$, we need to specify that there are four channels $D_1N(^2S_{\frac{1}{2}})$, $D_1N(^4D_{\frac{1}{2}})$, $D_2^*N(^4D_{\frac{1}{2}})$, and $D_2^*N(^6D_{\frac{1}{2}})$. If considering these four channels, with the obtained OPE potential we obtain the bound state solution of $J=1/2$ isoscalar $D_1N/D_2^*N$ state with the cutoff $
\Lambda$ around 1 GeV, where the dominant channel is $D_1N(^2S_{\frac{1}{2}})$ and the remaining three channels have very small probabilities, which show that this is a typical $D_1N$ bound state.
Our numerical work also shows that there does not exist the 
$D_1N/D_2^*N$ molecular state with $J=1/2$ and $I=1$ since we cannot find its bound state solution. 

For the $D_1N/D_2^*N$ systems with $J=3/2$ and $I=0,1$, six channels are considered, which are $D_1N(^4S_{\frac{3}{2}})$, $D_1N(^2D_{\frac{3}{2}})$, $D_1N(^4D_{\frac{3}{2}})$, $D_2^*N(^4S_{\frac{3}{2}})$, $D_2^*N(^4D_{\frac{3}{2}})$, and $D_2^*N(^6D_{\frac{3}{2}})$. With the cutoff $\Lambda$ around 1 GeV, the bound state solution for $D_1N/D_2^*N$ systems with $J=3/2$ and $I=0$ can be obtained, which indicates that there exist isoscalar $D_1N/D_2^*N$ systems with $J=3/2$. Since its dominant channel is $D_2^*N(^4S_{\frac{3}{2}})$, it is a typical $D_2^*N$ molecular state. 

In addition, we can find the bound state solution for $D_1N/D_2^*N$ system with $J=3/2$ and $I=1$, which has  a dominant $D_1N(^4S_{\frac{3}{2}})$ channel. However, the corresponding cutoff is around 3 GeV, which is far from the usual value of a cutoff. Thus, our result does not favor the existence of a $D_1N/D_2^*N$ molecular state with $J=3/2$ and $I=1$ if we strictly require $\Lambda$ to be around 1 GeV. 

The predicted two isoscalar $D_1N/D_2^*N$ molecular states with $J=1/1$ and $3/2$ have the same decay mode as that of $D_1N[0(\frac{1}{2})^+]$ and $D_2^*N[0(\frac{3}{2})^+]$, respectively.

\subsubsection{Case II}

When $\Lambda$ is taken to be around 2 GeV, the bound state solutions of the $D_1N/D_2^*N$ system with $J=1/2$ and $I=0,1$ appear, where the dominant contribution is from the $D_1N(^2S_{\frac{1}{2}})$ channel for both systems. 

For the $D_1N/D_2^*N$ system with $J=3/2$ and $I=0,1$, we can find the bound state solutions for the $D_1N/D_2^*N$ states with $J=3/2$ and $I=0,1$, where the values of the cutoff $\Lambda$ are around 1 GeV and 3 GeV for isoscalar and isovector $D_1N/D_2^*N$ states with $J=3/2$, respectively. The dominant channels of isoscalar and isovector $D_1N/D_2^*N$ systems with $J=3/2$ are $D_1N(^4S_{\frac{3}{2}})$ and $D_2^*N(^4S_{\frac{3}{2}})$, respectively. 

The above results show that there at least exists a isoscalar molecular state $D_1N/D_2^*N$ with $J=3/2$, which is a typical $D_1N$ molecular state.

Besides these discussions of the possible decay channels of the predicted molecular states presented in Secs. \ref{22} and \ref{11}, we also provide the information of the decay channels of other discussed $D_1N/D_2^*N$ systems with different quantum numbers (see Table \ref{tabIV} for more details). 

We notice the experimental progress on the observations of $\Lambda_c(2940)$, $\Sigma_c(2800)$, and $X(3250)$, all of which are from the $B$ meson decays. Thus, searchs for these predicted hadronic molecular states can be performed at Belle, LHCb, and forthcoming BelleII through the $B$ meson weak decays. In addition, the present study of the exotic molecular states composed of nucleon and $P$-wave charmed meson can be extended to explore the exotic
nuclei containing the $D_1(2420)$ or $D_2^*(2460)$ meson, which can be accessed at J-PARC and GSI by using
antiproton beams with targets of nuclei.

\section{Discussion and conclusion}

Experiments have resulted in big progress on the observations of new hadronic states over the past decade, which has inspired extensive discussions on how to explain these novel phenomena. Among different theoretical explanations, introducing the exotic state assignment to these new observed hadronic states is very popular. A peculiarity of the observed hadronic states is that the masses of these hadronic states are near the corresponding threshold of a hadron pair, which is also the reason why the molecular state explanation is often suggested.

Just introduced in Sec. \ref{sec1}, $\Lambda_c(2940)$ \cite{Aubert:2006sp}, $\Sigma_c(2800)$ \cite{Abe:2006rz}, and $X(3250)$ \cite{Lees:2012kc} have a common property, i.e., each of them is close to the corresponding threshold of a  charmed meson and a nucleon. Previously, $\Lambda_c(2940)$ as the $D^*N$ molecular state, $\Sigma_c(2800)$ as the $DN$ molecular state, and $X(3250)$ as the $D_0^*(2400)N$ were proposed and investigated in the literature.

If there are molecular states composed of a charmed meson and a nucleon, we predict that the exotic molecular states with a nucleon and a $P$-wave charmed meson in the $T$ doublet should also exist, which makes this molecular state family become more complete. For determining whether these molecular states exist, in this work we study the interaction of a nucleon with a charmed meson in the $T$ doublet through the OPE model, where the S-D mixing and the coupled channel effect are considered in our investigation.

Our numerical results indicate that a nucleon and a $P$-wave charmed meson with $J^P=0^+$ or $J^P=1^+$ in the $T$ doublet
can form bound states, where the detailed results can be found in Sec. \ref{sec3}. Thus, we suggest that our experimental colleagues to search for these exotic molecular states near the $D_1(2420)N$ and $D_2^*(2460)N$ thresholds, where Belle, LHCb, and the forthcoming Belle II are suitable platforms to carry out the experimental exploration of these exotic molecular states with a nucleon and a $P$-wave charmed meson.

\section*{Acknowledgments}
This project is supported
by the National Natural Science Foundation of China under
Grants No. 11222547, No. 11175073, No. 11035006, and No. 11311120054,
the Ministry of Education of China (SRFDP under Grant No. 20120211110002 and the Fundamental Research Funds for the Central
Universities), the Fok Ying-Tong Education
Foundation (Grant No. 131006), and RFBR under Grant No. 13-02-91154.


\end{document}